# Charge Transport in Electronic Devices Printed with Inks of Quasi-1D van der Waals Materials

Saba Baraghani,[1,2] Jehad Abourahma,[3] Zahra Barani,[1] Amirmahdi Mohammadzadeh,[1] Sriharsha Sudhindra,[1] Alexey Lipatov,[3] Alexander Sinitskii,[3,4] Fariborz Kargar,[1,2,*] and Alexander A. Balandin[1,2,*]

[1]Nano-Device Laboratory and Phonon Optimized Engineered Materials Center, Department of Electrical and Computer Engineering, University of California, Riverside, California 92521 USA

[2]Department of Chemical and Environmental Engineering, University of California, Riverside, California 92521 USA

[3]Department of Chemistry, University of Nebraska, Lincoln, Nebraska 68588 USA

[4]Nebraska Center for Materials and Nanoscience, University of Nebraska, Lincoln, Nebraska 68588 USA

[*] Corresponding authors: fkargar@ece.ucr.edu ; balandin@ece.ucr.edu ; web-site: http://balandingroup.ucr.edu/

## ABSTRACT

We report on fabrication and characterization of electronic devices printed with inks of quasi-1D van der Waals materials. The quasi-1D van der Waals materials are characterized by 1D motifs in their crystal structure, which allows for their exfoliation into bundles of atomic chains. The ink was prepared by the liquid-phase exfoliation of crystals of TiS$_3$ semiconductor into quasi-1D nanoribbons dispersed in a mixture of ethanol and ethylene glycol. The temperature dependent electrical measurements indicate that electron transport in the printed devices is dominated by the electron hopping mechanisms. The low-frequency electronic noise in the printed devices is of $1/f^\beta$-type with $\beta \sim 1$ near room temperature ($f$ is the frequency). The abrupt changes in the temperature dependence of the noise spectral density and $\gamma$ parameter can be indicative of the phase transition in individual TiS$_3$ nanoribbons as well as modifications in the hopping transport regime. The obtained results attest to the potential of quasi-1D van der Waals materials for applications in printed electronics.

- **INTRODUCTION**

Recent years have witnessed a boost in printed electronics research and development as the technique facilitates mass production of electronic devices with lower cost and processing requirements [1,2]. The approach also enables manufacturing of large-scale and flexible devices by expanding the choices of substrates from conventional silicon to flexible surfaces such as paper and textile [3–5]. Inks with proper thermophysical properties are the crucial component of the printing industry. Different functional materials have been developed and used as the ingredients for the inks employed by various printing techniques. The printed electronic devices can find applications in various areas. Flexible radio frequency tags,[6] wearable electronics,[4] organic light emitting diodes,[7] and organic solar cells[8] are just a few examples of such applications. Despite the recent advancements, however, the list of available materials as ingredients for the inks is limited. Little is known about the nature of the charge transport and electronic noise characteristics in printed electronic devices.

Some of the previously demonstrated inks are based on metal nanoparticles, such as Ag,[9,10] Au,[11] and Cu,[12] carbon allotropes, such as graphene[13] and carbon nanotubes[14], and nanowires, such as ZnO, dispersed in suitable solvents.[15] Most recently the attention has turned to the layered quasi-two-dimensional (2D) van der Waals (vdW) materials, such as transition metal dichalcogenides (TMDs), including $MoS_2$,[16,17] $MoSe_2$,[1] $WS_2$,[18] and $Bi_2Te_3$.[19] The class of quasi-2D vdW materials is promising for printing owing to a relative ease of ink development. It is known that TMD materials have weak vdW bonding between structural units allowing for their exfoliation into flexible quasi-2D layers. The inks of quasi-2D vdW materials can be prepared by the liquid-phase exfoliation (LPE) process and dispersion of the exfoliated flakes in proper solvents. Additionally, TMDs exhibit tunable electronic and mechanical properties, which increase their value as ingredients for various functional inks.[20–25]

The class of layered vdW materials is not limited to quasi-2D materials only. Most recently, there was a rapid emergence of interest to vdW materials with quasi-one-dimensional (1D) crystal structure.[26,27] These materials are quasi-1D in the sense that they have strong covalent bonds along the atomic chains, and vdW bonds or substantially weaker covalent bonds in directions perpendicular to the chains [28–30]. Transition metal trichalcogenides (TMTs), with chemical formula of $MX_3$, where M is a transition metal and X is a chalcogen, are a prominent group of quasi-1D vdW materials [28–31]. Examples of materials from this group include $TiS_3$,[32,33] $TaSe_3$,[28] $NbS_3$,[34] $ZrS_3$,[32,33,35,36] $ZrTe_3$,[37] and their solid solutions.[32] Unlike TMDs, which exfoliate into quasi-2D atomic planes, TMT crystals exfoliate into needle-like quasi-1D structures with high aspect ratios. [24,28,31,38] Motivated by the recent developments in the quasi-1D vdW material synthesis and exfoliation, we demonstrate the feasibility of the use of such materials in the inks for printed electronics. The high aspect ratio and flexibility of the exfoliated bundles of quasi-1D atomic chains offer specific advantages: possibly better connectivity of the individual flakes, smaller loading fractions required to achieve electrical conduction, and potentially more narrow printed lines. [39,40]

The intrinsic properties of quasi-1D vdW metals and semiconductors can add to the unique ink functionalities. The exfoliated quasi-1D TMTs have demonstrated exceptional electrical properties. For example, the bundles of $TiS_3$, $TaSe_3$ and $ZrTe_3$ have shown exceptionally high breakdown current densities of ~1.7 MA/cm$^2$, ~10 MA/cm$^2$, and ~100 MA/cm$^2$, respectively [41] For the present study, we selected $TiS_3$.[42–50] It is an n-type semiconductor with a bandgap of ~1 eV at room temperature (RT).[43,45,49–51] The material undergoes a metal – insulator transition at temperature $T_M \sim 220$ K and exhibits metal-like properties at higher temperatures.[52] Bulk $TiS_3$ whiskers were shown to have the RT Hall mobility of about 30 cm$^2$V$^{-1}$s$^{-1}$,[46] and a comparable exciton mobility of about 50 cm$^2$V$^{-1}$s$^{-1}$.[53] Theoretically, it has been suggested that a single quasi-1D monolayer of $TiS_3$ can have a mobility of ~ 10,000 cm$^2$V$^{-1}$s$^{-1}$, which is higher than that of quasi-2D $MoS_2$.[54] The experimentally observed mobilities in $TiS_3$ are considerably lower than the theoretical prediction, likely due to the polar-optical phonon scattering.[52,55] The few-layer $TiS_3$-based field-effect transistors (FETs) revealed mobilities of about 20-40 cm$^2$V$^{-1}$s$^{-1}$ and ON/OFF ratios of > 10$^3$;[24] these experimental values are comparable or higher than those in FETs based on few-layer $MoS_2$, which are ~10 to 20 cm$^2$V$^{-1}$s$^{-1}$.[56,57] A report on the LPE and drop casting of $TiS_3$[58] suggested that this material might be suitable for ink printing if proper solvent is found and printing parameters are optimized.

The crystal structure of monoclinic TiS$_3$ with the $P2_1/m$ space group is presented in Figure 1 (a). Blue and yellow spheres represent the Ti and S atoms, respectively. The figure shows that TiS$_3$ has a highly anisotropic structure, in which quasi-1D chains of TiS$_3$ prisms are covalently bonded along the *b*-axis. These weakly interacting chains are assembled into vdW-stacked quasi-2D layers parallel to the *ab* plane of the crystal structure. A previous theoretical study has shown that the cleavage energies required for breaking weak interactions between the quasi-2D layers separated by the vdW gaps (Figure 1 (a)) and between the quasi-1D chains within the layers are both comparable to the cleavage energy of graphene layers in graphite.[31] Because of this, a cleavage of a TiS$_3$ crystal may realistically proceed along several different planes between the quasi-1D chains (such as (001), (100), (101), etc.), resulting in a formation of high-aspect-ratio exfoliated nanoribbons.[31]

▪ **EXPERIMENTAL SECTION**

**Materials:** TiS$_3$ crystals were synthesized by the direct reaction between metallic titanium and a sulfur vapor in a vacuum-sealed quartz ampule at 550 °C, as described in our previous works.[24,53] Figure 1 (b) shows an optical photograph of the as-grown TiS$_3$ crystals. Because of their highly anisotropic crystal structure (Figure 1 (a)), these crystals are needle-shaped, with their long axes corresponding to the crystallographic *b* direction of the quasi-1D TiS$_3$ chains. The LPE process was performed on these TiS$_3$ crystals in ethanol, resulting in cleavage of the bundles of the atomic chains, which appeared as needle-like nanoribbons.

**Characterization:** The material quality after the exfoliation processing steps have been confirmed with Raman spectroscopy (Renishaw InVia). The measurements were performed in the backscattering configuration under visible red laser ($\lambda = 633$ nm). The excitation power on the surface was kept at ~2 mW. Figure 1(c) presents Raman spectrum of exfoliated TiS$_3$ bundles at RT. The monoclinic structure of TiS$_3$ has $C_{2h}$ point symmetry with irreducible representation of $\Gamma = 8A_g + 4B_g + 8B_u + 4A_u$. Four dominant Raman peaks identified at 174, 297, 367, and 556 cm$^{-1}$ belong to $A_g$ vibrational symmetries.[30,31,59] The peak at 174 cm$^{-1}$ is associated with the rigid chain vibrations.[59] The two peaks at 297 cm$^{-1}$ and 367 cm$^{-1}$ are related to the internal out-of-plane vibrations involving each monolayer. The peak at 556 cm$^{-1}$ is attributed to S – S diatomic motions.[59] The three low-intensity peaks at 163, 276, and 401 cm$^{-1}$ that shoulder to the dominant

Raman peaks belong to $B_g$ vibrational symmetries. [60,61] The Raman spectra confirm the material quality after the exfoliation processing steps.

To further confirm the high-quality and crystallinity of the solution-exfoliated TiS$_3$ nanoribbons, we studied them by transmission electron microscopy (TEM) (FEI Tecnai Orisis). TEM image of a representative TiS$_3$ nanoribbon is shown in Figure 1 (d). A selected area electron diffraction (SAED) pattern recorded on this nanoribbon (see the inset in Figure 1 (d)) confirms that it is highly crystalline. According to the indexing of the diffraction spots in this SAED pattern, the observed view corresponds to the *ab* plane of TiS$_3$. The long axis of the crystal represents to the crystallographic *b* direction of the quasi-1D chains, which is in agreement with the most expected exfoliation scenario.[31] A high-resolution TEM image in Figure 1 (e), which was recorded for the same nanoribbon, confirms that this is the *ab* plane, because the observed interplanar distances perfectly match the *a* and *b* crystal structure parameters of TiS$_3$ (*a* = 0.4948 nm, *b* = 0.3379 nm, *c* = 0.8748 nm, and the cant angle *β* = 97.62°; see the crystallographic analysis in ref. 24).

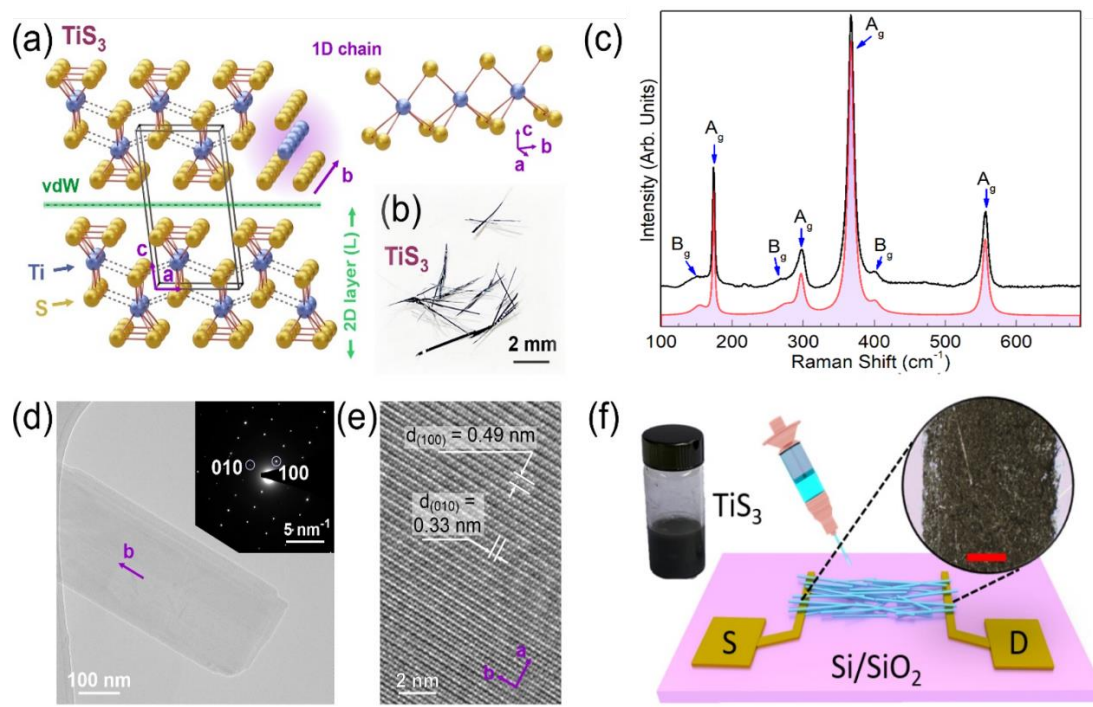

**Figure 1. Crystal structure, sample preparation and characterization.** (a) Schematic of the monoclinic crystal structure of TiS$_3$ from two viewpoints. The blue and yellow spheres represent the Ti and S atoms, respectively. The parallelogram demonstrates the unit cell of TiS$_3$. The side view in the right panel exhibits the quasi-1D nature of the atomic chains. (b) Optical photograph of the TiS$_3$ crystals used in this study. (c) Raman spectrum of exfoliated TiS$_3$ (black curve) at room temperature. The red curve shows the cumulative fitting of the experimental data by individual Gaussian functions. (d) TEM image of a representative solution-exfoliated TiS$_3$ nanoribbon. A SAED pattern recorded on this crystal is shown in the inset. (e) High-resolution TEM image of the same TiS$_3$ nanoribbon as in panel (d). (f) Schematic of the printing process of TiS$_3$ devices on top of gold contacts made with electron beam lithography. The image in the dashed circle is the optical image of the actual TiS$_3$ device channel. Note the absence of the "coffee-ring" effect in the channel confirming that the material is distributed evenly. The scale bar is 200 μm. The vial contains the liquid-phase exfoliated TiS$_3$ ink.

**Device Preparation:** A schematic of the printing procedure and device are shown in Figure 1 (f). The electrodes were fabricated by e-beam lithography followed by the lift-off of Ti/Au (20-nm/200-nm) deposited by e-beam evaporation. The TiS$_3$ channel was printed by a 3D printer (Hyrel 30M) on top of the electrodes. The printed channel has the dimensions of 500 μm×4 μm×6 μm (L×W×t). The thermophysical properties of the ink and proper selection of the injecting nozzle size play a crucial role in the ink droplet formation, even spreading of the ink on the substrate, and subsequent drying. The nonuniform deposition of the material in printing often happens as a result of the "coffee-ring" effect that should be avoided by adjusting the concentration of the ingredients in the ink and surface modification of the substrate.[62] The droplet formation behavior is characterized by a dimensionless Z-number (inverse of Ohnesorge number)[63,64], $Z = \sqrt{\gamma \rho a}/\mu$, where $a$ is the printer's nozzle diameter [m], and $\gamma$, $\rho$, and $\mu$ are the surface tension [N m$^{-1}$], density [kg m$^{-3}$], and dynamic viscosity [Pa × s] of the ink, respectively. Generally, proper droplet formation and dispense occur in the range of $1 \leq Z \leq 14$. Otherwise, satellite droplet formation ($Z > 14$) or elongated ligaments ($Z < 1$) may deteriorate the accuracy of the printing by deposition of the ink on undesired areas.[63,65–67]. However, there are studies that report quality printings with $Z \geq 14$, especially for nanomaterial-based inks. For example, graphene-based inks with $Z \sim 24$ or polystyrene-nanoparticle-based inks with $Z \sim 21$ to $\sim 91$ have been successfully printed.[65,68–71] Therefore, the Z-number limits seems not to be strict for vdW materials, and inks with Z-number values close to the range guarantee proper printing quality.

To tune the Z-number for printing with quasi-1D vdW ink, we used a syringe with inner diameter of $a = 210$ μm. The exfoliated TiS$_3$ bundles of atomic chains were mixed in ethanol with ethylene glycol (EG), using the mixing proportion of 1:1 (vol%). The dynamic viscosity, surface tension, and the mass density of the ink were measured to be $\gamma \sim 3.3$ [Nm$^{-1}$], $\mu \sim 3.35 \times 10^{-3}$ [Pa × s], and $\rho \sim 949.6$ [kgm$^{-3}$], respectively. The details of the measurements can be found in the METHODS section and the Supplementary Information. The Z-number of the prepared ink was $Z \sim 24$. No satellite droplet formation or elongated ligaments were observed during the printing. Addition of EG helped to circumvent the "coffee-ring".[72] A faster evaporation rate of ethanol causes enrichment of the droplet contact line with EG, leading to a surface tension gradient in the droplet and uniform dispersion of the material while drying.[2,67] To augment the flow through the ink droplet the bed temperature of the printer was kept at 60 °C.

## ▪ RESULTS AND DISCUSSION

The electrical current-voltage (I-V) characteristics of the fabricated two-terminal devices were measured in the temperature range from 78 K to 350 K. The I-V results are presented in Figure 2 (a). The weakly non-linear I-V curves were attributed to formation of the Schottky barrier at the printed channel – metal contact interface. Interestingly, it was previously shown that Au forms an Ohmic contact with TiS$_3$,[73] provided that there is a clean ultrahigh-vacuum-enabled interface between the two materials resulting in strong Au-S interactions. The formation of Schottky barrier for these devices can be explained by the fabrication process, where the interfaces between the printed TiS$_3$ nanoribbons and the Ti/Au pads contain various surface adsorbates and residual solvent molecules. At low bias voltage, from -0.1 V to 0.1 V, the I-V characteristics can be considered approximately linear (see the inset to Figure 2 (a)). The resistivity of the printed TiS$_3$ channel was extracted from the linear I-V region as a function of temperature (see Figure 2 (b)). At RT, the resistivity of the printed devices was ~195 Ω×m. The resistivity values reported in the literature span a wide range from 0.02 Ωm to 200 Ω×m for individual exfoliate bundles of atomic chains, *i.e.* nanoribbons, and crystals with defects.[52,58,74] Note that our resistivity results include the contact resistance. The solvent residues can also affect the overall resistance of the channel. Given the disordered nature of the printed device, the extracted values of resistivity are reasonable and consistent with data reported for individual nanoribbons.[52]

As one can see in Figure 2 (b), the resistivity continuously decreases with increasing temperature. This is in contrast with the data on the electrical resistivity measurements of bulk and mechanically exfoliated few-layer TiS$_3$ devices reported over an extended temperature range of 4 K to 400 K.[43,46,47,52]. In the latter case, the resistivity decreased with the temperature increasing up to ~250 K, which is typical for semiconductor material. At temperatures above 250 K, TiS$_3$ exhibits metallic properties, *i.e.* the resistivity increases with the temperature rise.[52] In printed devices, however, one would expect to have a network of randomly arranged exfoliated TiS$_3$ atomic chain bundles with many bundle to bundles interfaces, defects, and impurities resulting from the LPE process and subsequent printing. In this case, one would expect the electron hopping transport to be the main mechanism of electron conduction. The electrically conductivity due to electron hopping has smaller values than the bulk band conduction, and it increases with increasing temperature. We now analyze the resistivity data in more details.

Several models have been proposed to describe the electron hopping in disordered material systems[75]. In the nearest neighbor hoping (NNH) model, the system is considered to have randomly distributed isoenergetic sites with a concentration of $N_0$ and the electron localization length of $\alpha$ so that $N_0\alpha \ll 1$. The electrical conduction is carried out by the charge carriers jumping between the nearest sites. The temperature dependence of the electrical conductivity in NNH mechanism, $\sigma_{NNH}$, is described by the equation $\sigma_{NNH} = \sigma_{0NNH}\exp(-E_{NNH}/k_BT)$, where $k_B$ is Boltzmann constant, $\sigma_{0NNH}$ and $E_{NNH}$ are the NNH conductivity constant and the NNH activation energy, respectively. This equation is similar to the equation for the electrical conductivity in the thermally activated band-conduction model, $\sigma = \sigma_0\exp(-E_a/k_BT)$, where $E_a$ is the thermal activation energy. The main difference in the formalisms is that $E_{NNH} < E_a$ so that generally NNH occurs at rather high temperature where the thermal activation energy is not sufficient to excite electrons to the conduction band but enough to excite them to the available spatially-separated energy sites between the conduction and valence bands. At the low-temperature limit, the electron hopping is dominated by the variable-range hoping (VRH) mechanism [75–79] where the upward transition of electrons between energetically distinct but spatially nearby states is less likely than the transition to states with energetically close but spatially farther away states. Our experimental data at low temperature can be better described with the Efros-Shklovskii VRH model [75,76,79] in which the hoping distance between the trap sites is not constant and carriers can hop between the levels closer to the Fermi level.[76] The temperature-dependent conductivity in this model is described by the equation $\sigma = \sigma_{0ES}\exp[-(T_{ES}/T)^{1/2}]$, where $\sigma_{0ES}$ and $T_{ES}$ are parameters that depend on the localization length and dielectric constant of the material.[79]

Figure 2 (c) shows the Arrhenius plot of the resistivity data in the entire examined temperature range.[76,80] The plot is divided into three regions shaded in yellow, blue, orange, and green colors. The blue and green regions correspond to the temperature ranges where the slope of the resistivity changes significantly. The change in the slope of the curve in blue region is a signature of transitioning from VRH (yellow) to NNH (orange) conduction in disordered materials.[77,81–84] The activation energy calculated for the NNH conduction is ~12.4 meV. As expected, the extracted value is significantly lower than the thermal activation energy reported for bulk TiS$_3$ crystals which is ~43 meV.[44,46,52] The data deviates from the NNH model at high temperature again (see green area in Figure 2 (c)). The reason for such a deviation needs a further investigation. Figure 2 (d) shows the resistivity data as a function of $T^{-1/2}$ in the temperature range from 78 K to 115 K where VRH conduction dominates. According to the Efros-Shklovskii VRH model, the plot of $\ln(\sigma)$ as a function of $T^{-1/2}$ must be linear. One can see an excellent agreement between our experimental data and the fitted linear regression in this temperature range.

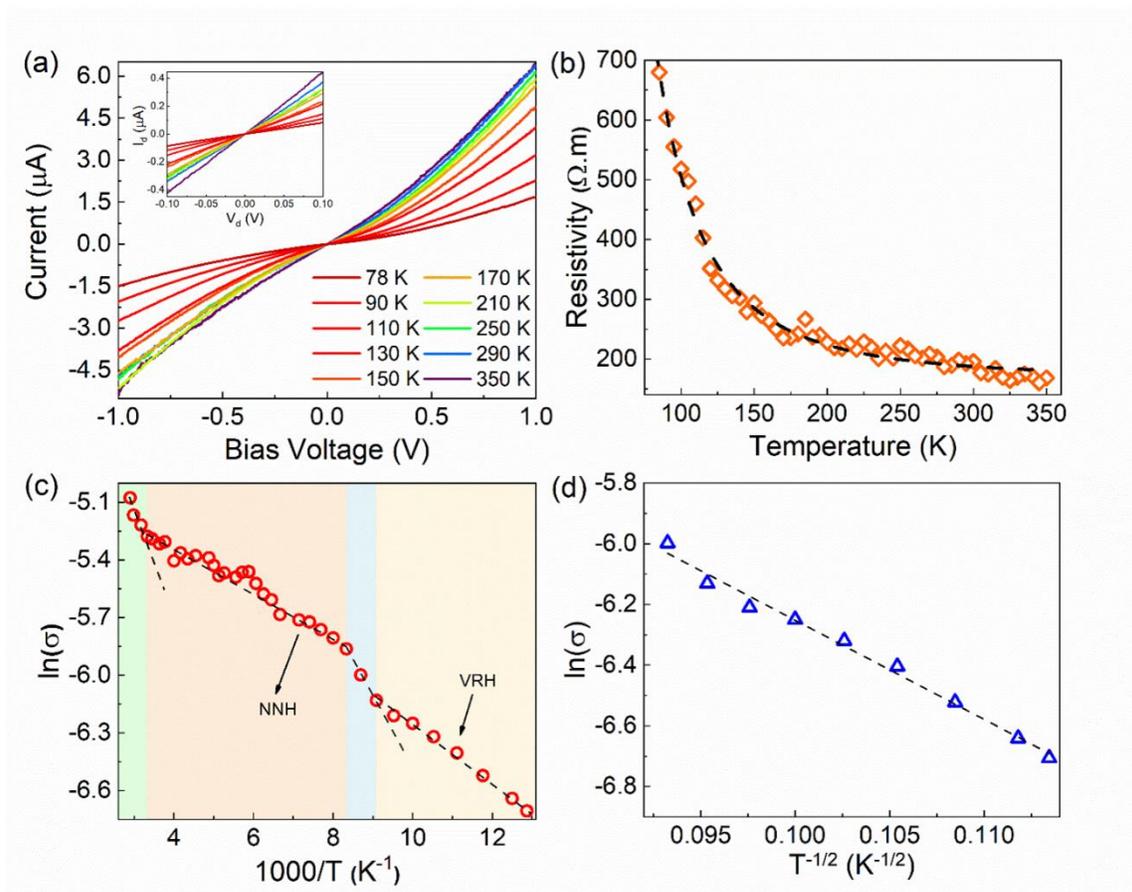

**Figure 2. Electron transport mechanism in the devices printed with 1D van der Waals ink.** (a) Current-voltage characteristics of the TiS$_3$ printed device as a function of temperature. The inset shows the linear I-V dependence at small bias voltages. At higher voltages, the I-V curves become non-linear. (b) Electrical resistivity of the printed TiS$_3$ channel as a function of temperature. The decrease in resistivity with temperature increase is consistent with the electron hopping transport mechanism. (c) Arrhenius plot of the electrical conductivity of the printed TiS$_3$ channel. Different shades of color in the plot indicates changes in conduction mechanism from nearest-neighbor hopping (NNH) to variable-range hopping (VRH). (d) Plot of $\ln(\sigma)$ versus $T^{-1/2}$ for the printed device channel at low tempratures. The experimental data agrees with the theoretical Efros - Shklovskii variable range hoping model (dashed line).

We used the low-frequency electronic noise spectroscopy to further elucidate the electron transport properties in our printed devices. The details of our experimental setup and measurement procedures have been reported elsewhere in the context of other material systems [83,85]. The analysis of the noise spectral density, its functional dependence on frequency, electric bias and temperature can provide a wealth of information on electron transport, particularly in material systems with high concentration of defects and impurities which act at the charge trapping sites. We have successfully used the electronic noise spectroscopy for monitoring phase transitions in materials which reveal strongly correlated phenomena [83,86–89] Typically, at the frequencies $f<100$ kHz, materials show the spectral noise density of $S(f) \sim 1/f^\beta$ type, with $\beta \sim 1$. In Figures 3 (a) and (b) we present the voltage-referred noise power spectral density, $S_v$, and the normalized current noise power spectral density, $S_I/I^2$, for the device channel printed with the quasi-1D $TiS_3$ ink as a function of frequency. The noise spectra were measured for several bias voltages at RT. One can see that the noise generally follows the $1/f$ trend, and it increases with the increase in bias voltage as expected [83,86,89]. There are some traces of Lorentzian-type bulges at frequencies above $f$=100 Hz. They can indicate a presence of certain defects or impurities with particularly high concentration that act as the trapping centers for the charge carriers contributing to the current conduction. If such a defect, with a characteristic time constant of the charge carrier trapping and de-trapping, dominates the current fluctuations, its contribution to noise spectrum appears as a Lorentzian bulge.[90]

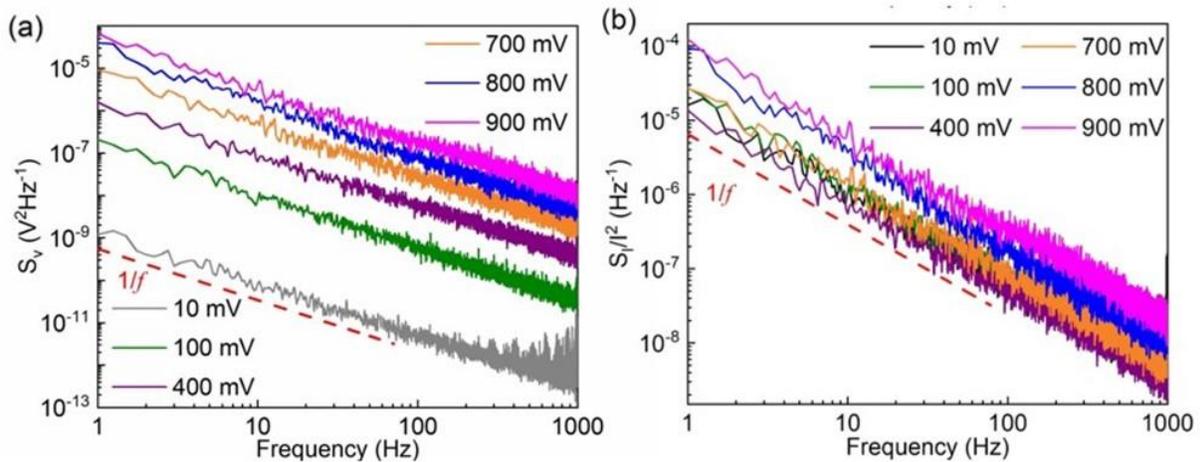

**Figure 3. Room-temperature electronic noise spectroscopy of the printed devices.** (a) Voltage-referred noise power spectral density, and (b) the normalized current noise power spectral density as the function of frequency at different applied bias voltages. The data in (a) and (b) follows $1/f$ noise dependency.

Figures 4 (a) and (b) show the voltage noise spectral density, $S_v$, and the normalized current spectral density, $S_I/I^2$, as a function of temperature. All measurements were carried out with the small applied bias of ~0.3 V to avoid Joule heating. The spectra in both plots follows the $1/f^\beta$. However, $\beta$ is no longer close to 1, and it reveals a rather strong functional dependence on temperature. Figure 4 (c) shows the extracted values of $\beta$ as a function of temperature in the range from 130 K to 350 K. The black dashed lines are eye guides only. The observed large deviation of $\beta$ from 1 can be related to the Lorentzian bulges (see Figure 4 (a) and (b)), which are more pronounced at low and high temperature limits. The changes in the noise spectra can be associated with the reported metal – insulator transition at temperature $T_M$~250 K [52] as well as the change in the electron hopping conductivity around T~320 K (see green region in Figure 2 (c)). Figure 4 (d) shows the normalized noise spectral density, $S_I/I^2$, as a function of temperature at the fixed frequency of 10 Hz. The noise level abruptly increases at the temperature of ~320 K. This supports the hypothesis of changes in the electron hopping transport mechanism as seen in resistivity data in Figure 2 (c) and $\gamma$ parameter dependence in Figure 4 (c).

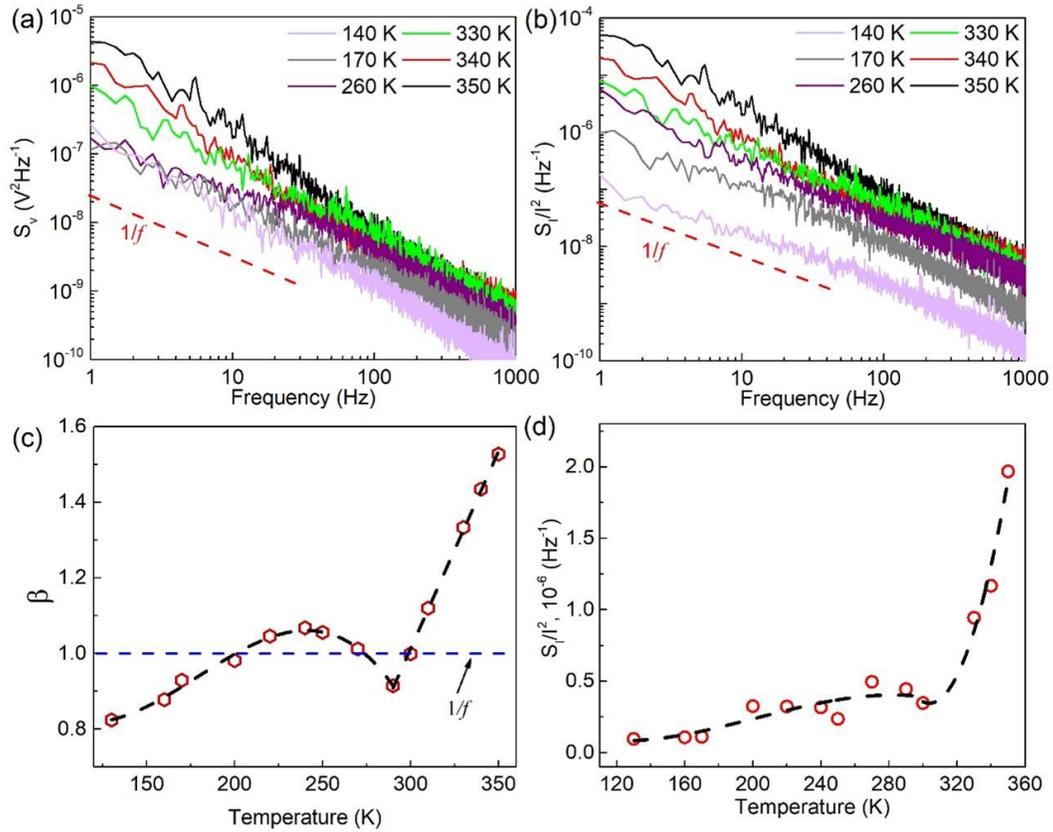

**Figure 4. Temperature-dependent low-frequency electronic noise characteristics of the printed devices.** (a) Low-frequency noise spectra of voltage fluctuations, $S_v$, as a function of frequency measured at different temperatures. (b) Normalized current noise spectral density as a function of frequency for various temperatures. (c) Extracted values of $\beta$ as a function of temperature. (d) Normalized noise spectral density versus temperature at the constant frequency of $f = 10$ kHz.

### ▪ CONCLUSIONS

In conclusions, we reported on printing electronic devices with inks of quasi-1D vdW materials. The ink was prepared by the LPE of small crystals of $TiS_3$ semiconductor into quasi-1D nanoribbons dispersed in a mixture of ethanol and ethylene glycol. The temperature dependent electrical measurements indicate that electron transport in the printed devices is dominated by the electron hopping mechanisms. The low-frequency electronic noise in the printed devices is of $1/f^{\beta}$-type with $\beta \sim 1$ near room temperature ($f$ is the frequency). The abrupt changes in the temperature dependence of the noise spectral density and $\gamma$ parameter can be indicative of the phase transition in individual TiS$_3$ nanoribbons as well as modifications in the hopping transport regime. The obtained results attest to the potential of quasi-1D vdW materials for applications in printed electronics. The developed printing process can also facilitate characterization of new quasi-1D van der Waals materials predicted by the machine learning studies, which are being currently synthesized.

### ▪ METHODS

**TiS$_3$ ink preparation and characterization:** 3 mL of 4 mg/mL of exfoliated TiS$_3$ in ethanol was mixed with 3 mL of ethylene glycol (EG) to give 6 mL of 2 mg/mL of TiS$_3$ in 1/1 vol% ethanol/ethylene glycol as the ink. TEM of the solution-exfoliated TiS$_3$ nanoribbons was performed using a FEI Tecnai Osiris scanning transmission electron microscope at the accelerating voltage of 200 kV. Ethylene glycol was added to the ink to increase the viscosity of the ink. This is to ensure proper dispensing and drying of the ink. The viscosity, surface tension, and density of the ink were measured to calculate the characteristic Z number. To measure the viscosity, a Cannon SimpleVIS viscometer was used. 0.5 μL of the ink was inserted in the device and the measured kinematic viscosity was displayed on the device screen. Then this value was divided by density to calculate the dynamic viscosity. A CSC Scientific DuNouy interfacial tensiometer was used to calculate the surface tension of the ink. To measure the surface tension, the device ring was inserted in the ink. Then, the ring was raised slowly until the film between the ink and the ink breaks, and the measured amount of surface tension can be read from the device dial. With the addition of EG to the ink, the viscosity and surface tension of the ink were measured to be 3.346 mPa.s and 33.1 mN m$^{-1}$. The density of the ink was calculated using the rule of density for mixtures. The calculated density of the ink was 949.6 kg.m$^{-3}$. With these values the calculated Z number of the ink is 24.28 for the nozzle with inner diameter of 210 μm used in this work.

**Printing devices with quasi-1D TiS$_3$ vdW materials:** The TiS$_3$ ink was used in a Hyrel 30M system 3D printer to print out the TiS$_3$ channel. A blunt-end needle syringe with gauge 27 was used with the printer. The bed temperature of the printed was kept at 60 °C during the whole course of printing process. 20 layers of material were printed on the substrate with the gold electrodes. The electrodes were fabricated by electron-beam lithography and lift-off of Ti/Au (20-/200-nm), deposited by electron-beam evaporation. The final TiS$_3$ channel had a channel length of 4 μm. The width and thickness of the channel were 500 μm and 6 μm, respectively.

**Electrical measurements and noise spectroscopy:** All current–voltage (I–V) characteristics and resistivities were measured in the cryogenic probe station (Lakeshore TTPX) with a semiconductor analyzer (Agilent B1500). The low-frequency noise experiments were conducted in the two-terminal device configuration. The noise spectra were measured with a dynamic signal analyzer (Stanford Research 785). A battery biasing circuit was used to apply bias voltage to the devices. This was done to minimize the noise at 60 Hz and the harmonics associated with it. The signal measured by the dynamic signal analyzer is the absolute voltage noise spectral density, $S_v$, of a parallel resistance network consisting of a load resistor ($R_L$) of 46 KΩ and the device under test with a resistance of $R_D$. The normalized current noise spectral density, $S_I/I^2$, was calculated $S_I/I^2 = S_V \times [(R_L + R_D)/(R_L \times R_D)]^2/(I^2 \times G)$ where G is the amplification of the low-noise amplifier.

■ **ASSOCITED CONTENT**

**Supporting Information**

The supporting information is available free of charge on the ACS Publication website at DOI: XXXX

> The supporting information includes a detailed description of the liquid phase exfoliation and material characterization, printing and device preparation, as well as electrical characterization of the device channel.

■ **AUTHOR INFORMATION**

**Corresponding Author**
* Email: balandin@ece.ucr.edu (A.B.B.).
* Email: fkargar@ece.ucr.edu.

**ORCID**

Saba Baraghani: 0000-0003-0256-0397

Zahra Barani Beiranvand: 0000-0002-4850-0675

Amirmahdi Mohammadzadeh: 0000-0002-6852-2126

Sriharsha Sudhindra: 0000-0001-5299-3962

Alexey Lipatov: 0000-0001-5043-1616

Fariborz Kargar: 0000-0003-2192-2023

Alexander A. Balandin: 0000-0002-9944-7894

**ACKNOWLEDGEMENTS**

The work at UC Riverside was supported, in part, by the National Science Foundation (NSF) program Designing Materials to Revolutionize and Engineer our Future (DMREF) via a project DMR-1921958 entitled "Collaborative Research: Data Driven Discovery of Synthesis Pathways and Distinguishing Electronic Phenomena of 1D van der Waals Bonded Solids". The work at UNL was supported by the NSF, through ECCS 1740136, as well as by the nCORE, which is a wholly owned subsidiary of the Semiconductor Research Corporation (SRC), through the Center on Antiferromagnetic Magneto-electric Memory and Logic (AMML), task No. 2760.002.

**CONTRIBUTIONS**

A.A.B. and F.K. conceived the idea, coordinated the project, contributed to experimental data analysis, and led the manuscript preparation; S.B. prepared the ink, printed the devices, conducted current-voltage and electronic noise measurements, and contributed to data analysis; J.A., A.L. and A.S. provided the $TiS_3$ crystals, characterized them by TEM, and contributed to the data analyses; A.M. contributed to electrode fabrication; Z.B. assisted with the material exfoliation and ink preparation and conducted Raman measurements; S.S. carried out optical profilometer; All authors contributed to writing the manuscript.

S. Baraghani et al., Charge Transport in Electronic Devices Printed with Inks of Quasi-1D van der Waals Materials (2021)## REFERENCES

(1) Rowley-Neale, S. J.; Foster, C. W.; Smith, G. C.; Brownson, D. A. C.; Banks, C. E. Mass-Producible 2D-MoSe2 Bulk Modified Screen-Printed Electrodes Provide Significant Electrocatalytic Performances towards the Hydrogen Evolution Reaction. *Sustain. Energy Fuels* **2017**, *1* (1), 74–83. https://doi.org/10.1039/C6SE00115G.

(2) Hu, G.; Yang, L.; Yang, Z.; Wang, Y.; Jin, X.; Dai, J.; Wu, Q.; Liu, S.; Zhu, X.; Wang, X.; Wu, T.-C.; Howe, R. C. T.; Albrow-Owen, T.; Ng, L. W. T.; Yang, Q.; Occhipinti, L. G.; Woodward, R. I.; Kelleher, E. J. R.; Sun, Z.; Huang, X.; Zhang, M.; Bain, C. D.; Hasan, T. A General Ink Formulation of 2D Crystals for Wafer-Scale Inkjet Printing. *Sci. Adv.* **2020**, *6* (33), eaba5029. https://doi.org/10.1126/sciadv.aba5029.

(3) Zschieschang, U.; Klauk, H. Organic Transistors on Paper: A Brief Review. *J. Mater. Chem. C* **2019**, *7* (19), 5522–5533. https://doi.org/10.1039/C9TC00793H.

(4) Bidoki, S.; McGorman, D.; Lewis, D.; Clark, M.; Horler, G.; Miles, R. E. Inkjet Printing of Conductive Patterns on Textile Fabrics. *AATCC Rev.* **2005**, *5*, 11–14.

(5) Carey, T.; Cacovich, S.; Divitini, G.; Ren, J.; Mansouri, A.; Kim, J. M.; Wang, C.; Ducati, C.; Sordan, R.; Torrisi, F. Fully Inkjet-Printed Two-Dimensional Material Field-Effect Heterojunctions for Wearable and Textile Electronics. *Nat. Commun.* **2017**, *8* (1), 1202. https://doi.org/10.1038/s41467-017-01210-2.

(6) Redinger, D.; Molesa, S.; Yin, S.; Farschi, R.; Subramanian, V. An Ink-Jet-Deposited Passive Component Process for RFID. *IEEE Trans. Electron Devices* **2004**, *51* (12), 1978–1983. https://doi.org/10.1109/TED.2004.838451.

(7) Park, S.-I.; Xiong, Y.; Kim, R.-H.; Elvikis, P.; Meitl, M.; Kim, D.-H.; Wu, J.; Yoon, J.; Yu, C.-J.; Liu, Z.; Huang, Y.; Hwang, K.; Ferreira, P.; Li, X.; Choquette, K.; Rogers, J. A. Printed Assemblies of Inorganic Light-Emitting Diodes for Deformable and Semitransparent Displays. *Science (80-. ).* **2009**, *325* (5943), 977 LP – 981. https://doi.org/10.1126/science.1175690.

(8) Ganesan, S.; Mehta, S.; Gupta, D. Fully Printed Organic Solar Cells – a Review of Techniques, Challenges and Their Solutions. *Opto-Electronics Rev.* **2019**, *27* (3), 298–
18 | P a g e